\documentclass[aps,pra,superscriptaddress,12pt,tightenlines,nofootinbib]{revtex4}
\usepackage{amsmath,amsthm,graphicx,amssymb,verbatim,listings}
\usepackage[usenames, dvipsnames]{color}
\usepackage[ pdftex, plainpages = false, pdfpagelabels,
                 pdfpagelayout = useoutlines,
                 bookmarks,
                 bookmarksopen = true,
                 bookmarksnumbered = true,
                 breaklinks = true,
                 linktocpage=all,
                 pagebackref=false,
                 colorlinks = true,
                 linkcolor = BrickRed,
                 urlcolor  = blue,
                 citecolor = BrickRed,
                 anchorcolor = green,
                 hyperindex = true,
                 hyperfigures
                 ]{hyperref}
\usepackage{xifthen} 

\newcommand{\tr}{\hbox{tr}}

\newcommand{\ket}[1]{{\ensuremath{\left| #1 \right\rangle}}}
\newcommand{\bra}[1]{{\ensuremath{\left\langle #1 \right|}}}
\newcommand{\braket}[2]{{\ensuremath{\left\langle #1 \middle| #2
      \right\rangle}}}

\newcommand{\arxiv}[2][]{\ifthenelse{\isempty{#1}}{\href{http://arxiv.org/abs/#2}{{\tt arXiv:\allowbreak{}#2}}} {\href{http://arxiv.org/abs/#2}{{\tt arXiv:\allowbreak{}#2 [#1]}}}}

\newcommand{\booktitle}{\textsl}
\newcommand{\hrefdoi}[2]{\href{https://dx.doi.org/#1}{#2}}

\begin{document}

\title{SIC-POVMs and Compatibility among Quantum States}

\author{Blake C.\ Stacey}
\affiliation{Department of Physics,
  University of Massachusetts Boston, 100 Morrissey Blvd.,
  Boston, MA 02125, United States}

\begin{abstract}
An unexpected connection exists between compatibility criteria for
quantum states and symmetric informationally complete POVMs.
Beginning with Caves, Fuchs and Schack's ``Conditions for
compatibility of quantum state assignments'' [Phys.\ Rev.\ A\ {\bf 66}
  (2002), 062111], I show that a qutrit SIC-POVM studied in other
contexts enjoys additional interesting properties.  Compatibility
criteria provide a new way to understand the relationship between
SIC-POVMs and mutually unbiased bases, as calculations in the SIC
representation of quantum states make clear.  This, in turn,
illuminates the resources necessary for magic-state quantum
computation, and why hidden-variable models fail to capture the
vitality of quantum mechanics.
\end{abstract}

\maketitle

This article presents an unforeseen connection between two subjects
originally studied for separate reasons by the Quantum Bayesians, or
to use the more recent and specific term, QBists~\cite{Voldemort,RMP}.
One of these topics originates in the paper ``Conditions for
compatibility of quantum state
assignments''~\cite{CavesFuchsSchack2002} by Caves, Fuchs and Schack
(CFS).  Refining CFS's treatment reveals an unexpected link between
the concept of compatibility and other constructions of quantum
information theory.

We begin by taking up the question of what it means for probability
distributions to be compatible with one another.  Consider a
thoroughly classical scenario, in which Alice and Bob are gambling on
the outcome of a coin toss.  Both Alice and Bob are certain the toss
is rigged, but Alice is convinced that the outcome will be heads,
while Bob is equally steadfast in maintaining that it will be tails.
That is, $p_A(H) = 1$ and $p_A(T) = 0$, while $p_B(H) = 0$ and $p_B(T)
= 1$.  No matter which way the coin lands, one party will be
disappointed---or, depending on the stakes, bankrupt.

We can equally well present a single-user version of this scenario.
Imagine that Alice, and only Alice, is gambling on the coin flip, and
that conditional on some other information, she will choose to do so
in accordance with either the probability distribution
\begin{equation}
p_A(H) = 1,\qquad p_A(T) = 0;
\end{equation}
or the alternative,
\begin{equation}
p_A'(H) = 0,\qquad p_A'(T) = 1.
\end{equation}

For example, Alice may be confident that the toss will be rigged and
that she can deduce which way it will be rigged once she can observe
the handedness of the coin-tosser.  The incompatibility between $p_A$
and $p_A'$ is then relevant to Alice, regardless of the presence or
absence of other players.

Having formulated the scenario in single-user terms, we can develop a
quantum analogue; a single-user statement avoids the conceptual
problem of whether an event occurring for an agent Alice could, in the
quantum setting, itself be an event for any other agent~\cite{RMP}.
The quantum version of this kind of incompatibility is a condition on
pairs of quantum states.  Take two density matrices $\rho_A$ and
$\rho_A'$.  If there exists a measurement $\{E_i\}$ such that
\begin{equation}
\sum_i \tr (\rho_A E_i) \tr (\rho_A' E_i) = 0,
\end{equation}
then $\rho_A$ and $\rho_A'$ are called \emph{post-Peierls
  incompatible}~\cite{CavesFuchsSchack2002}.  If two states are
post-Peierls (PP) compatible, then for \emph{all} experiments, there
is at least one outcome for which both states yield nonzero
probability via the Born rule.  We can naturally extend this criterion
to sets of more than two quantum states.  In general, a set of~$N$
states is PP incompatible when, for some experiment $\{E_i\}$,
\begin{equation}
\sum_i \prod_{a=1}^N \tr (\rho^{(a)} E_i) = 0.
\end{equation}

A von Neumann measurement is a Positive Operator Valued Measure (POVM)
with $d$ possible outcomes, specified by a set of~$d$ orthonormal
vectors in Hilbert space.  In terms of projection operators, each von
Neumann measurement comprises $d$ one-dimensional orthogonal
projectors.  We abbreviate this phrase as ODOP.  Therefore,
compatibility with respect to von Neumann measurements is known as
\emph{PP-ODOP} compatibility \cite{CavesFuchsSchack2002}.  When we
formulate an exact criterion for PP-ODOP compatibility of
\emph{qutrit} pure states, \emph{i.e.}, pure states in~$d = 3$, we
find something interesting.

In the next two sections, we will examine PP-ODOP compatibility in
more detail and find a connection to another topic of much interest to
the QBist research program.  For completeness, we note that
compatibility criteria have also recently entered the quantum
foundations discourse through a different route.  They play a key role
in discussions of whether quantum states can be treated as encoding
information about the values of hidden variables~\cite{PBR,
  Schlosshauer14}.  In this paper, we disregard the issue of hidden
variables and treat quantum states as directly specifying the
probabilities of experiment outcomes.

\section{Three Pure States in Dimension Three}

It is possible to have triplets of states which are PP-ODOP
incompatible when taken all together, even though they are compatible
when taken in pairs.  CFS provide a specific example, which for
convenience we reproduce here.  Pick an orthonormal basis $\{\ket{0},
\ket{1}, \ket{2}\}$, and consider the following three possible states
which can be ascribed to a qutrit:
\begin{align}
\ket{\psi} &= \frac{1}{\sqrt{2}}\left(\ket{1} + \ket{2}\right),
 \nonumber\\
\ket{\psi'} &= \frac{1}{\sqrt{2}}\left(\ket{2} + \ket{0}\right),
\label{eq:example-set}\\
\ket{\psi''} &= \frac{1}{\sqrt{2}}\left(\ket{0} + \ket{1}\right).
 \nonumber
\end{align}

Now, Alice the graduate student performs a von Neumann experiment in
the computational basis of $\{\ket{0}, \ket{1}, \ket{2}\}$.  Whatever
result she experiences, there is a state assignment in the set $\{
\ket{\psi}, \ket{\psi'}, \ket{\psi''}\}$ according to which that
experience is an event of probability zero.

By pursuing this argument, we will have the opportunity, incidentally,
to correct two mathematical errors in CFS.  Both are slight, but one
is a missed opportunity, as it introduced an inconsistency into CFS's
calculations and obscured the connection which we shall examine here.

A general set of three pure states $\{\ket{\psi}, \ket{\psi'},
\ket{\psi''}\}$ is PP-ODOP incompatible at the tertiary level if some
orthonormal basis $\{\ket{0}, \ket{1}, \ket{2}\}$ exists such that
\begin{align}
\ket{\psi} &=
 e^{i\chi}\left(\cos\theta\ket{1}
               + e^{i\phi}\sin\theta\ket{2}\right),
 \nonumber\\
\ket{\psi'} &=
 e^{i\chi'}\left(\cos\theta'\ket{2}
               + e^{i\phi'}\sin\theta'\ket{0}\right),\\
\ket{\psi''} &=
 e^{i\chi''}\left(\cos\theta''\ket{0}
               + e^{i\phi''}\sin\theta''\ket{1}\right).
 \nonumber
\end{align}

The three $\theta$ angles are restricted to the interval $(0, \pi/2)$,
while the $\chi$ and $\phi$ angles must all lie in the interval $[0,
  2\pi)$.  Taking the inner products picks out one basis vector per
  pair:
\begin{align}
\braket{\psi}{\psi'}
 &= e^{i(\chi'-\chi)} e^{-i\phi}
    \sin\theta\cos\theta',\nonumber\\
\braket{\psi'}{\psi''}
 &= e^{i(\chi''-\chi')} e^{-i\phi'}
    \sin\theta'\cos\theta'',\\
\braket{\psi''}{\psi}
 &= e^{i(\chi-\chi'')} e^{-i\phi''}
    \sin\theta''\cos\theta. \nonumber
\end{align}

This is the first of the two errata mentioned above: because the
$\phi$-dependent phase factors come from the bra vectors rather than
the ket vectors, the sign in the exponential should be negative,
instead of positive as written in CFS's equation (19).  The sign error
cancels in the next step, which is to multiply these quantities by
their complex conjugates, yielding
\begin{align}
\left|\braket{\psi}{\psi'}\right|^2
 &= \sin^2\theta\cos^2\theta', \nonumber\\
\left|\braket{\psi'}{\psi''}\right|^2
 &= \sin^2\theta'\cos^2\theta'',
 \label{eq:PP-ODOP-trig} \\
\left|\braket{\psi''}{\psi}\right|^2
 &= \sin^2\theta''\cos^2\theta.\nonumber
\end{align}

After some algebra, one can prove from Eq.~(\ref{eq:PP-ODOP-trig})
that three pure states in~$d = 3$ are PP-ODOP incompatible if and only
if
\begin{align}
\left|\braket{\psi}{\psi'}\right|^2
 + \left|\braket{\psi'}{\psi''}\right|^2
 + \left|\braket{\psi''}{\psi}\right|^2 &< 1
\label{eq:PP-ODOP-sym1},\\
\left(\left|\braket{\psi}{\psi'}\right|^2
 + \left|\braket{\psi'}{\psi''}\right|^2
 + \left|\braket{\psi''}{\psi}\right|^2 - 1\right)^2
 &\geq 4\left|\braket{\psi}{\psi'}\right|^2
     \left|\braket{\psi'}{\psi''}\right|^2
     \left|\braket{\psi''}{\psi}\right|^2.
\label{eq:PP-ODOP-sym2}
\end{align}

This is the second glitch in CFS: the latter inequality should be
$\geq$ rather than $>$, as written in CFS's Formula (28).  This is the
more consequential error, since if the inequality is strict, then the
example of three-party PP-incompatible states which CFS provide---our
triplet Eq.~(\ref{eq:example-set})---is not consistent with CFS's
compatibility criterion.

\section{Qutrit SIC POVMs}

The second inequality has a more intricate structure than the first.
What happens when we try to {\it saturate\/} it?  Suppose we require
that the three squared overlaps all have the same value, $x$.  Then,
saturating the second inequality implies that $x$ satisfies the cubic
equation
\begin{equation}
4x^3 - 9x^2 + 6x - 1 = 0.
\end{equation}

This cubic polynomial has a zero at $x = \frac{1}{4}$, and a double
zero at~$x = 1$, which is disallowed by the requirement that the
states are nonidentical.  

How many states can we push simultaneously to the edge of PP-ODOP
incompatibility in this way?  That is, how many states in qutrit state
space can we find such that for any two of them,
\begin{equation}
\left|\braket{\psi_i}{\psi_j}\right|^2 = \frac{1 + 3\delta_{ij}}{4}\ ?
\end{equation}

This is the problem of finding the {\it maximal set of equiangular
  lines\/} in three-dimensional complex vector space.

Beginning with the question of compatibility among probability
distributions, we have arrived at {\it Symmetric Informationally
  Complete POVMs}~\cite{RMP, Zauner99, Renes04, Scott10,
  ConicalDesigns}.  This term is abbreviated to SIC-POVM, or just to
SIC (pronounced like ``seek'').  A SIC for a $d$-dimensional Hilbert
space is a set of $d^2$ operators $\{E_i = \frac{1}{d} \Pi_i\}$ where
the rank-one projection operators $\{\Pi_i\}$ satisfy
\begin{equation}
\tr (\Pi_k \Pi_l) = \frac{d\delta_{kl} + 1}{d + 1}.
\label{eq:sic-trace}
\end{equation}

It is known that SICs are maximal in this regard, \emph{i.e.,} no more
than $d^2$ operators can simultaneously satisfy
Eq.~(\ref{eq:sic-trace}).  For qutrits, this means that a set of
states such that any three saturate the edge of PP-ODOP
incompatibility can contain at most \emph{nine} states.

We have shown that any triple of states chosen from a qutrit SIC will
be PP-incompatible.  One might expect that a large number of different
von Neumann measurements would be required to cover all the possible
choices of triples, perhaps comparable to the number of triples
themselves.  Surprisingly, this is not the case; our toolbox can be
much more economical.  Take $\omega=e^{2\pi i/3}$, and construct the
set of states $\{\ket{\psi_j}\}$ given by the columns of the following matrix:
\begin{equation}
\frac{1}{\sqrt{2}}
\left(
\begin{array}{ccccccccc}
0 & -1 & 1 & 0 & -1 & 1 & 0 & -1 & 1 \\
1 & 0 & -1 & \omega & 0 & -\omega & \omega^2 & 0
 & -\omega^2 \\
-1 & 1 & 0 & -\omega^2 & \omega^2 & 0
 & -\omega & \omega & 0
\end{array}
\right).
\label{eq:ME-SIC}
\end{equation}


The set of nine states $\{\ket{\psi_i}\}$ forms a SIC known as the
\emph{Hesse SIC}~\cite{Dang2013}.  For the Hesse SIC, a set of {\it
  four\/} orthogonal bases is sufficient to reveal the
PP-incompatibility of all possible triples.  Moreover, the requisite
bases have an interesting property: they are \emph{mutually unbiased}
with respect to each other.  In general, two bases are mutually
unbiased if, for any vector $\ket{\psi}$ in one basis and any vector
$\ket{\phi}$ in the other, $|\braket{\psi}{\phi}|^2 = 1/d$.  Any set
of three states drawn from the SIC will be revealed as PP-incompatible by
a measurement in one or more of the Mutually Unbiased Bases (MUB).  We
construct each basis vector by finding a state orthogonal to three of
the SIC states.  Specifically, each basis vector corresponds to an
element in a Steiner triple system~\cite{Cameron2002} of order 9,
which we build by cyclically tracing all the horizontal, vertical and
diagonal lines in the array
\begin{equation}
\begin{array}{ccc}
0 & 1 & 2\\
3 & 4 & 5\\
6 & 7 & 8
\end{array}
; \qquad\hbox{that is to say, }\qquad
S(9) =
\begin{array}{ccc}
(012) & (345) & (678) \\
(036) & (147) & (258) \\
(048) & (156) & (237) \\
(057) & (138) & (246)
\end{array}.
\label{eq:steiner}
\end{equation}

Each possible value of the index $i$ occurs in exactly four entries of
$S(9)$, and each possible \emph{pair} of index values occurs exactly
once.  It is easiest to see the meaning of this construction using the
SIC representation of quantum states.  Any quantum state, pure or
mixed, is equivalent to a probability distribution over the outcomes
of an informationally complete measurement, and the qutrit SIC of
Eq.~(\ref{eq:ME-SIC}) furnishes such a measurement.  An arbitrary
qutrit density matrix $\rho$ can be decomposed as
\begin{equation}
\rho
 = \sum_{i=0}^{8} \left(4 p(i) - \frac{1}{3}\right) \Pi_i
 = 4 \sum_{i=0}^{8} p(i) \Pi_i - I,
\label{eq:rho-to-sic}
\end{equation}
where $\Pi_i = \ket{\psi_i}\bra{\psi_i}$ and $p(i)$ is the Born-rule
probability
\begin{equation}
p(i) = \frac{1}{3} \tr (\rho\Pi_i).
\end{equation}

To construct the state orthogonal to SIC vectors $\ket{\psi_i}$,
$\ket{\psi_j}$ and $\ket{\psi_k}$, we simply write a probability
vector $\textbf{p}$ which is zero in entries $i$, $j$ and $k$, and
$\frac{1}{6}$ everywhere else.

To see why this construction yields a complete set of MUB, we use the
fact that the Hilbert--Schmidt inner product of density matrices is
just an affine transformation of the Euclidean inner product of the
corresponding probability distributions~\cite{RMP, Fuchs2014}.  For
qutrit states,
\begin{equation}
\tr\,\rho\sigma = 12\textbf{p}\cdot\textbf{q} - 1.
\end{equation}
This means that if $\rho$ and $\sigma$ are orthogonal, then
$\textbf{p}\cdot\textbf{q} = 1/12$.

With this, we can see that the vectors orthogonal to $(012)$ and
$(345)$, for example, must be orthogonal to each other, because when
we take the dot product of their SIC representations, we only have
three nonzero contributions.  If instead we take the vectors
orthogonal to $(012)$ and $(036)$, say, a zero in one vector coincides
with a zero in the other, the dot product can come out larger.
These within-row and between-rows relationships hold generally.  Each
row corresponds to a set of three mutually orthogonal vectors, and
when we take vectors from two different rows, we always get the same
nonzero overlap: the Hilbert space inner product of their density
matrices is always $\frac{1}{3}$.

We have fashioned a \emph{complete set of mutually unbiased bases.}
Starting with the SIC Eq.~(\ref{eq:ME-SIC}), we constructed 12
pure states which fall into four sets of three.  Each set of three,
corresponding to a row in our table, is an orthonormal basis.  When we
take the Hilbert--Schmidt inner product of a state from one basis with
a state from another, we get $1/3$ every time.  This is the
requirement for two bases to be mutually unbiased in~$d=3$ (in older
language, the observables associated with any two such bases are
complementary \cite{schwinger1960}).  Furthermore, the largest number
of MUB that can exist in~$d$ dimensions is $d+1$~\cite{appleby2009},
and we constructed four.  From now on, we will refer to these $d(d+1)
= 12$ vectors as \emph{MUB states} for short.  While the relation
between qutrit SIC and MUB states has been known for some
time~\cite{Bengtsson10, Dang2013}, the convenience of the SIC
representation has so far not been appreciated.

Given any three distinct elements from the SIC set, a measurement in
at least one of the MUB will reveal PP-ODOP incompatibility among
those three states.  For example, say we pick the SIC elements
$\ket{\psi_0}$, $\ket{\psi_1}$ and $\ket{\psi_4}$.  Then, we measure in
the basis given by the vectors orthogonal to the fourth row: $(057)$,
$(138)$, $(246)$.  Each possible outcome of the experiment conflicts
with one of the three given states: the first with the state
ascription $\ket{\psi_0}$, the second with the ascription
$\ket{\psi_1}$ and the third with $\ket{\psi_4}$.  Therefore, we have PP-ODOP
incompatibility at the ternary level, while of course any two distinct
states in the SIC are \emph{pairwise} compatible, having an
inner-product-squared of~$1/(d+1) = 1/4$.

Note that the second error in CFS, writing $>$ instead of $\geq$ in
the condition Eq.~(\ref{eq:PP-ODOP-sym2}), mistakenly implies that
triplets of SIC states are PP-ODOP compatible, though barely so.  This
is clearly incorrect, as we can see by testing $\ket{\psi_0}$,
$\ket{\psi_1}$ and $\ket{\psi_2}$ in the computational basis.

\section{Additional Properties of the Hesse SIC and Associated MUB}

In order to be a SIC representation of a \emph{pure} quantum state, a
probability distribution $p(i)$ must satisfy two
conditions~\cite{RMP}.  The first is quadratic:
\begin{equation}
\sum_i p(i)^2 = \frac{2}{d(d+1)};
\label{eq:purity1}
\end{equation}
and the second is cubic, or ``QBic'':
\begin{equation}
\sum_{ijk} \hbox{Re}[\tr(\Pi_i\Pi_j\Pi_k)] p(i)p(j)p(k)
 = \frac{d+7}{(d+1)^3}.
\label{eq:purity2}
\end{equation}

The quadratic condition has an interesting interpretation that
provides a handy mnemonic for it~\cite{stacey-thesis}.

Whenever we have a probability distribution, we can compute indices
that summarize its properties.  One well-known example is the Shannon
entropy, also designated as the Shannon information or Shannon index,
which reflects the extent to which a probability distribution is
``spread out.''  The Shannon index is maximized for a uniform
distribution, and it attains its minimum value of zero when the
distribution is a delta function.  Another way to quantify the spread
of a probability distribution is an \emph{effective number.}

Imagine that we have an urn full of marbles, and we presume that when
we draw a marble from the urn, no choice is preferred over any other.
If the urn contains $N$ marbles, our probability of obtaining any
individual one of them is $1/N$.  However, what if our probability
distribution is not uniform, as it would be if we thought the drawing
was rigged in some way?  In that case, we can label the marbles with
the integers from~1 to~$N$, and we say that our probability for
obtaining the one labeled $i$ is $p(i)$.

We draw one marble, replace it and repeat the drawing.  What is the
probability that we will draw the same marble both times?  Let the
result of the first drawing be $j$.  Then, our probability for
obtaining that marble again is $p(j)$, and to find the overall
probability for drawing doubles, we average over all the choices
of~$j$:
\begin{equation}
\hbox{Prob}(\hbox{doubles}) = \sum_j p^2(j).
\end{equation}
For a uniform distribution, this is
\begin{equation}
\sum_j p^2(j) = \sum_j \left(\frac{1}{N}\right)^2
 = N \left(\frac{1}{N}\right)^2
 = \frac{1}{N}.
\end{equation}
That is, if all draws are equally probable, then the probability of a
coincidence is the reciprocal of the population size.  Turning this
around, we can say that whatever our probabilities for the different
draws, the effective size of the population is
\begin{equation}
N_{\rm eff} = \left[ \sum_j p^2(j) \right]^{-1}.
\end{equation}

Amusingly, assigning a pure state to a quantum system
means that the \emph{effective number} of possible outcomes for a SIC
experiment that one is willing to contemplate is a simple
combinatorial quantity:
\begin{equation}
N_{\rm eff} = \binom{d+1}{2}.
\end{equation}
This provides a way to remember the value on the right-hand side of
the quadratic constraint, Eq.~(\ref{eq:purity1}).  Another way to
think of this is that when all SIC outcomes are judged as
equiprobable, that is to say $p(i) = \frac{1}{d^2}$, the effective
number of experimental outcomes is the total number which comprise the
SIC: $N_{\rm eff} = d^2$.  Thus, if we focus on the quadratic
constraint, ascribing a pure state means neglecting $\binom{d}{2}$
possible outcomes of a SIC experiment.  Entertainingly, this is also
the best known upper bound on the number of entries which can be zero
in a quantum-state assignment ${\bf p}$ \cite{Fuchs2014}.  This is not
a coincidence: we can deduce that bound by starting with the
normalization of ${\bf p}$ and squaring to find
\begin{equation}
\left( \sum_i p(i) \right)^2 = 1.
\end{equation}

We then apply the Cauchy--Schwarz inequality to find, writing $n_0$
for the number of zero-valued entries in ${\bf p}$,
\begin{equation}
(d^2 - n_0) \sum_{{\rm nonzero}} p(i)^2 \geq 
 \left( \sum_{{\rm nonzero}} p(i) \right)^2 = 1.
\end{equation}

We see the inverse of the effective number appearing on the left-hand
side.  Consequently,
\begin{equation}
n_0 \leq d^2 - N_{\rm eff},
\label{eq:weak-zeros-bound}
\end{equation}
and from Eq.~(\ref{eq:purity1}) we know the right-hand side
equals $d(d-1)/2$, as advertised. In earlier work~\cite{appleby2011},
it was conjectured that this bound might be improved, and that the
true upper bound on the number of zeros was actually $d$.  Note that
in $d = 3$, this is equivalent to the bound in
Eq.~(\ref{eq:weak-zeros-bound}).  However, using the so-called
\emph{Hoggar SIC} in dimension $d = 8$, we can construct states that
saturate the bound in Eq.~(\ref{eq:weak-zeros-bound}), containing
exactly 28 zeros.  This follows readily from the recent results of
Szymusiak and S\l{}omczy\'nski~\cite{Szymusiak2015}.  Therefore,
Eq.~(\ref{eq:weak-zeros-bound}) is actually the tightest bound
possible in general.

Using the Hesse SIC to define a representation for qutrit state space,
the QBic condition can be simplified to
\begin{equation}
\sum_i p(i)^3 - 3 \sum_{(ijk) \in S(9)} p(i) p(j) p(k) = 0.
\label{eq:qbic-tabia}
\end{equation}
Here, $S(9)$ is the Steiner triple system defined above.  This is a
consequence~\cite{Tabia12, TabiaAppleby13} of the fact that for the
Hesse SIC, the triple products
\begin{equation}
C_{jkl} = \hbox{Re}\,\tr(\Pi_j \Pi_k \Pi_l)
\end{equation}
take a particularly simple form.  Note that if all three indices are
equal, then $\tr(\Pi_j \Pi_k \Pi_l)$ reduces to $\tr(\Pi_j)$, which is
unity.  Likewise, if two of the three indices are equal, then the
value of the triple product follows from the definition of a SIC,
Eq.~(\ref{eq:sic-trace}).  The nontrivial case is when all three
indices differ.

All known SICs have a \emph{group covariance} property: they can be
generated by starting with a single vector (the so-called
``fiducial''), and applying the elements of a group to that vector to
create all the others.  In all cases but one, that group is the
\emph{Weyl--Heisenberg group}~\cite{Zhu2014}, which is defined from
the two generators
\begin{equation}
 X\ket{j} = \ket{j+1} \hbox{\ (modulo $d$)},\  Z\ket{j} = \omega^j \ket{j},
  \hbox{ where } \omega = e^{2i\pi / d}.
\label{eq:X-and-Z}
\end{equation}

Combinations of $X$ and $Z$, together with phase factors, yield the
Weyl--Heisenberg displacement operators:
\begin{equation}
\sigma_{a,b} = \left(-e^{i\pi / d}\right)^{ab} X^a Z^b.
\end{equation}

Starting with the fiducial vector $\ket{\psi_0}$, we create the other
vectors in the SIC by applying $\sigma_{a,b}$, with $a,b \in
\{0,\ldots,d-1\}$.  Because $X^d = Z^d = I$, it is convenient to
visualize the Weyl--Heisenberg displacement operators as living at the
points of a $d \times d$ grid.  In $d = 3$, this grid is just the
square array from Eq.~(\ref{eq:steiner}).

Group covariance alone tells us something about the triple products:
if acting with a group element $g$ on the projectors transforms them
as
\begin{equation}
\Pi_i \rightarrow g\Pi_i g^\dag = \Pi_{i'},
\end{equation}
then the cyclic property of the trace implies
\begin{equation}
C_{ijk} = C_{i'j'k'}.
\end{equation}

In dimension $d = 3$, we have a grid of nine points that we can carve
up into four different ``striations'' (horizontal, vertical and two
diagonal).  Each striation is a set of three parallel lines,
corresponding to three vectors in an orthonormal basis.  The
Weyl--Heisenberg operators are horizontal and vertical shifts of this
grid.  These shifts map one line in a striation into another.  Any
triple product corresponds to a set of three points in the grid.
Therefore, if a triple product belongs to one of the four striations,
we can transform it into any other triple product in that striation,
by applying a Weyl--Heisenberg operator and possibly permuting
indices.  Consequently, triple products are constant on striations.
The Hesse SIC has the additional nice property that triple products
are constant from one set of parallel lines to another.
The upshot of this is that for the Hesse SIC, we can find all the
nontrivial triple products entirely geometrically.  If $j$, $k$ and
$l$ are three \emph{collinear} points, then
\begin{equation}
C_{jkl} = -\frac{1}{8}.
\end{equation}
Otherwise, if $j$, $k$ and $l$ are distinct but noncollinear,
\begin{equation}
C_{jkl} = \frac{1}{16}.
\end{equation}
The fact that the triple products follow this geometrical rule
is what allows us to reduce the QBic equation (\ref{eq:purity2}) to
the simpler form of Eq.~(\ref{eq:qbic-tabia}).

If we have a probability distribution, we can compute the Shannon
entropy of it.  We can, therefore, ask which pure states maximize or
minimize the Shannon entropy of their SIC representations.  In
particular, if we try to minimize the Shannon entropy of~$\textbf{p}$
under the constraint that its ``effective number'' is
\begin{equation}
\left[\sum_i p(i)^2\right]^{-1} = \frac{d(d+1)}{2},
\end{equation}
then we find that $p(i)$ must be 0 in exactly three entries, and
uniformly $1/6$ elsewhere (this is pointed out, in slightly
different language, by Szymusiak and
S\l{}omczy\'nski~\cite{Szymusiak2015}).

How many such vectors are valid quantum states?  We must check them
against the QBic equation (\ref{eq:qbic-tabia}).  For any vector of
this form,
\begin{equation}
\sum_i p(i)^3 = 6 \left(\frac{1}{6}\right)^3 = \frac{1}{36}.
\end{equation}
Therefore, we must have
\begin{equation}
\sum_{(ijk) \in S(9)} p(i) p(j) p(k) = \frac{1}{108}.
\end{equation}

Suppose that we fill in one line of our $3 \times 3$ grid with zeros.
If $i$, $j$ and $k$ are the points on this line, or on any line that
intersects with it, then $p(i)p(j)p(k)$ will evaluate to zero.
Exactly two lines will correspond to nonzero products, namely, the two
lines parallel to the one we filled with zeros.  Therefore,
\begin{equation}
\sum_{(ijk) \in S(9)} p(i) p(j) p(k) = 2 \left(\frac{1}{6}\right)^3
 = \frac{2}{216} = \frac{1}{108}.
\end{equation}

It follows that the states we seek are the twelve states made by
filling one line in the $3 \times 3$ grid with zeros and inserting
$1/6$ elsewhere.  These twelve states fall naturally into four sets of
three, corresponding to the four rows of our table.  Each row is
derived from one way of carving the grid into three parallel lines.

Having a complete set of MUB, we can define a discrete Wigner
function~\cite{Wootters1987}.  Like a SIC representation, a Wigner
representation is a way of writing a quantum state as a list of real
numbers.  Unlike the SIC representation, Wigner functions for quantum
states can have negative values, and thus are called
``quasi-probability distributions''~\cite{Zhu2016}.  It is easiest to
define a Wigner quasi-probability function when the Hilbert-space
dimension is a prime number or a power of a prime.

Wootters~\cite{Wootters1987} showed that one can construct a set of
\emph{phase-space point operators} that live in a $d \times d$ grid
and enjoy the following properties.  First, each of the $d^2$
operators is Hermitian and has unit trace:
\begin{equation}
\tr A_j = 1.
\end{equation}
Second, they are orthogonal to one another:
\begin{equation}
\tr A_j A_k = d\delta_{jk}.
\end{equation}
Third, if we carve the grid into a set of parallel lines
$\{\lambda\}$, then
\begin{equation}
P_\lambda = \frac{1}{d} \sum_{j \in \lambda} A_j
\end{equation}
defines a set $\{P_\lambda\}$ of mutually orthogonal projection
operators, the sum of which is the identity.  For any density matrix
$\rho$, we have
\begin{equation}
\rho = \sum_j W(j) A_j,\ W(j) = \frac{1}{d} \tr(\rho A_j).
\end{equation}

Wootters' discrete Wigner function is closely related to
MUB~\cite{Gibbons2004}.  Each of the $d+1$ sets of parallel lines
corresponds to a basis, and these $d+1$ bases (containing $d$ states
each) are mutually unbiased with respect to one another.  Summing the
Wigner function along a line yields the probability of obtaining the
outcome corresponding to that state when performing a measurement in
the basis to which that line belongs.

For qutrit states, the Wigner functions will be quasi-probability
distributions over nine points,
\begin{equation}
\sum_{i=0}^8 W(i) = 1,
\end{equation}
where the individual $W(i)$ can go negative.  By summing $W(i)$ over a
line $(jkl)$, we get the probability $q_{jkl}$ of obtaining the
outcome corresponding to that vector if we perform a measurement on
that basis.  Alternatively, if we know these probabilities, we can
solve for the Wigner function at any point:
\begin{equation}
W(i) = \frac{1}{3} \left[ \sum_{\lambda: i \in \lambda} q_\lambda -
  1\right].
\end{equation}

Call $\textbf{s}_{ijk}$ the SIC representation of the MUB vector that
has zeroes in positions $i$, $j$ and $k$.  Then, for example,
\begin{align}
q_{012} &= 12 \textbf{p} \cdot \textbf{s}_{012} - 1 \\
 &= 1 - 2(p(0) + p(1) + p(2)).
\end{align}

If we add up all the probabilities involving point 0, we find that
$p(0)$ occurs four times in the sum, and all the other SIC
probabilities $p(j)$ occur once:
\begin{equation}
q_{012} + q_{036} + q_{045} + q_{057}
 = 4 - 2(3p(0) + p(0) + p(1) + \cdots + p(8))
 = 2 - 6p(0).
\end{equation}
Therefore,
\begin{equation}
W(0) = \frac{1}{3}(2 - 6p(0) - 1) = \frac{1}{3} - 2p(0).
\end{equation}
The argument works analogously for all the points in the grid, and so
we arrive at the relation
\begin{equation}
W(i) = \frac{1}{3} - 2p(i).
\label{eq:Wigner-SIC}
\end{equation}

The MUB states affiliated with the Hesse SIC have simple SIC
representations, as we have seen.  Their Wigner representations are
easily found using Eq.~(\ref{eq:Wigner-SIC}).  Let $W_{jkl}$ be the
quasi-probability function for the MUB state that is orthogonal to the
SIC vectors $\ket{\psi_j}$, $\ket{\psi_k}$ and $\ket{\psi_l}$.  Then,
\begin{equation}
W_{jkl}(i) = \left\{\begin{array}{cc}
 \frac{1}{3}, & i \in \{j, k, l\} \\
 0, & \hbox{otherwise}
 \end{array}\right. .
\end{equation}

Any two MUB vectors belonging to different bases intersect in one
point on the $3 \times 3$ grid. It is instructive to compare the
Wigner functions for these MUB vectors to the twelve states of maximal
knowledge in Spekkens' quasi-classical model of a three-level
system~\cite{spekkens2014}.

Moreover, it also follows from Eq.~(\ref{eq:Wigner-SIC}) that no
entry in the Wigner quasi-probability can go more negative than it
does for the Hesse SIC states themselves.  Generally, if
$\ket{\psi_k}$ is a SIC state used to define a SIC representation of
quantum state space, and if we turn that representation upon
$\ket{\psi_k}$ itself, then we find it corresponds to the probability
distribution:
\begin{equation}
e_k(i) = \frac{1}{d(d+1)} + \frac{1}{d+1} \delta_{ik}.
\end{equation}

In dimension $d = 3$, this is a probability vector that contains $1/3$
in the $k$th position and $1/12$ everywhere else.
Using Eq.~(\ref{eq:Wigner-SIC}) to turn this into a Wigner
quasi-probability, we find that
\begin{equation}
W_k(i) = \left\{\begin{array}{cc}
 -\frac{1}{3}, & i = k \\
 \frac{1}{6}, & i \neq k
\end{array}\right. .
\end{equation}

No entry in a SIC representation of a quantum state can exceed $1/d$.
This follows~\cite{appleby2011} from the quadratic condition,
Eq.~(\ref{eq:purity1}).  Therefore, 
\begin{equation}
W(i) \geq -\frac{1}{3}.
\end{equation}

In fact, it is known~\cite{Veitch14} that the sum of \emph{all} the
negative entries in a qutrit Wigner quasi-probability function cannot
exceed $1/3$ in magnitude.  The SIC states themselves pack as much
negativity into a single entry as a state can have.  This is why the
Hesse SIC states are among the \emph{maximally magic resources} for
quantum computation~\cite{Veitch14}.

The Hesse SIC states are nine in number, and in the terminology
of~\cite{Veitch14} are the ``Strange states.''  The other maximally
magic states---that is, the other states for which the sum of the
negative entries has maximal magnitude---are designated the ``Norrell
states.''  They are 36 in number, and they spread an equal share of
negativity across two entries of the Wigner representation.  To
illustrate, we write the Wigner representation of a Hesse SIC state as
a $3 \times 3$ grid:
\begin{equation}
\left(\begin{array}{ccc}
-1/3 & 1/6 & 1/6 \\
1/6 & 1/6 & 1/6 \\
1/6 & 1/6 & 1/6
\end{array}
\right).
\end{equation}

There are obviously nine ways to position the $-1/3$.  If we instead pick
two elements to be equal negative values, then we can form a state
like
\begin{equation}
\left(\begin{array}{ccc}
-1/6 & -1/6 & 1/3 \\
1/6 & 1/6 & 1/6 \\
1/6 & 1/6 & 1/6
\end{array}
\right).
\end{equation}

There are $\binom{9}{2} = 36$ ways to do this: First, we pick a
striation (horizontal, vertical, left diagonal or right diagonal).
Then, we pick a line within that striation (for which we have three
choices).  Finally, we select which element in that line we will set
to~$1/3$ (again, three choices).  Each set of nine states derived from
a striation is, in fact, a SIC.  Thus, the 36 Norrell states comprise
four separate SICs.  In the language of group theory, they are a
\emph{Clifford orbit,} where the \emph{Clifford group} is defined as
the stabilizer of the Weyl--Heisenberg group.

\begin{figure}[h]
\begin{center}
\includegraphics[width=10cm]{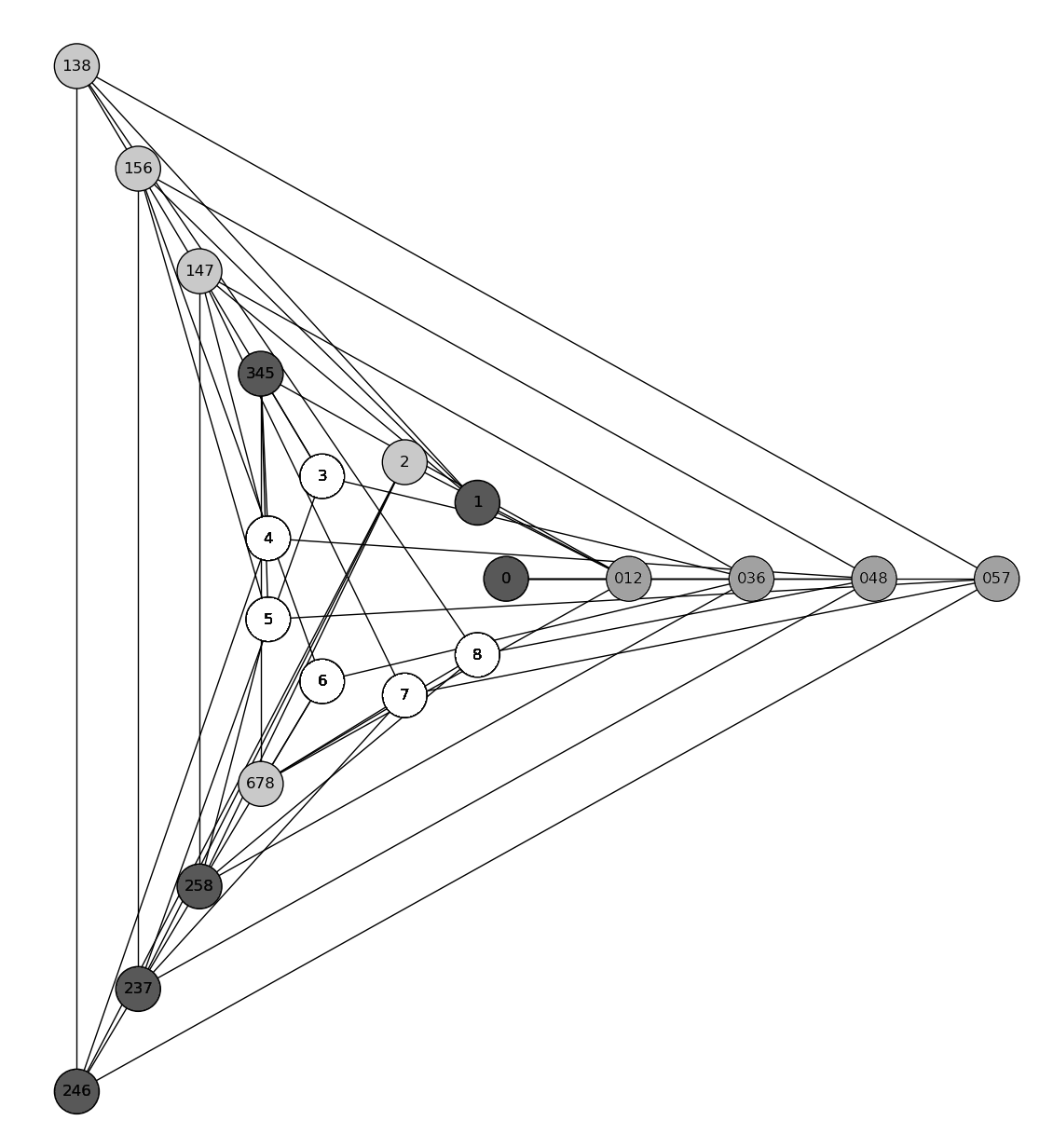}
\end{center}
\caption{\label{fig:hesse-graph} One possible presentation of the
  orthogonality graph for the Hesse SIC states and their associated
  MUB.  Each vertex stands for a quantum state, and vertices are
  linked by an edge if those states are orthogonal.  The nine circles
  near the center, labeled by single-digit numbers, denote the SIC
  states.  The other vertices, labeled by three-digit sequences, stand
  for the MUB states.  Each of the four nested triangles is an
  orthonormal basis, \emph{e.g.,} $\{\textbf{s}_{012},
  \textbf{s}_{345}, \textbf{s}_{678}\}$.  The graph is colored such
  that no two adjacent vertices share the same color.}
\end{figure}

We have seen how we can start with the Hesse SIC and construct four
MUB by minimizing the Shannon entropy of the SIC representation,
subject to the pure-state constraints.  Now, take the nine SIC vectors
and twelve MUB vectors, and represent each vector by a vertex of a
graph.  Connect two vertices with an edge if the corresponding states
are orthogonal.  The resulting graph has \emph{chromatic number}
4. That is, one needs at least four different colors of paint in order
to color all the vertices in such a way that no two adjacent vertices
share the same color.  We illustrate this in
Figure~\ref{fig:hesse-graph}.  Because the chromatic number exceeds
the dimension of the Hilbert space, this set of 21 vectors meets
Cabello's necessary condition for demonstrating ``state-independent
contextuality''~\cite{Cabello2011, Cabello2015}.

We now unpack the meaning of this statement. 

The issue at stake is \emph{whether quantum mechanics could be an
  approximation to some deeper theory of physics that is classical in
  character.}  Could it be that the randomness we find in quantum
phenomena might be explained away as due to our ignorance of more
fundamental degrees of freedom?  A saga of theorems, conceived by
Bell, Kochen, Specker and others, argues against
this~\cite{mermin1993}.  Quantum theory, they tell us, is incompatible
with the idea that quantum uncertainty is a result of our ignorance
about ``hidden variables'' contained within the systems we study
(so-called ``noncontextual hidden variables'').

It is not \emph{a priori} obvious exactly which phenomena truly tap
into this failure of classicality.  Many ``quantum'' effects can be
emulated in models that are essentially classical.  The list is
remarkably long, in fact, and includes teleportation, key
distribution, the no-cloning and no-broadcasting theorems, coherent
superpositions turning to incoherent mixtures by becoming entangled
with the environment, ``quantum discord'' and many
more~\cite{spekkens2007, vanenk2007, bartlett2012, spekkens2014,
  combes2015}.  However, the Bell--Kochen--Specker results take us out of
that regime.

We can think of the Kochen--Specker theorem (and its more modern
descendents) in the following way.  Suppose that we have some physical
system, and we list a series of ``questions'' that we might ask it.
Each question is some physical experiment that yields a quantitative
result.  For simplicity, we can assume that all these experimental
tests are binary, yielding either 0 or 1.  Prior to choosing a test
and carrying it out, one can have expectations about what will
transpire should one choose a particular test and perform it.  If we
assume that the behavior of the system is governed by some hidden
internal degrees of freedom that are independent of the test one might
elect to make, then this assumption will constrain the expectations
that one might have for the experimental outcomes.  The predictions
for different experiments will be tied together in a certain way---one
which quantum phenomena can violate.

When can a set of questions demonstrate this effect?  In quantum
theory, we can represent a binary question by a projection operator.
A set of projection operators defines an orthogonality graph.  A
necessary condition for a set of projectors to be able to reveal the
failure of the hidden-variable hypothesis is that the chromatic number
of their orthogonality graph exceed the dimension of the system's
Hilbert space~\cite{Cabello2011, Cabello2015}.  As explained above,
our set of SIC and MUB vectors meets this criterion.

Cabello's criterion is necessary but not sufficient.  However, the
Hesse SIC states and the MUB vectors we derived from them are, in
fact, sufficient to demonstrate nonclassicality.  A Kochen--Specker
theorem that demonstrates this explicitly has been worked
out~\cite{Bengtsson10}.

When taken together, the Hesse SIC and its affiliated MUB comprise a
set of questions, for which the statistics of the answers mesh
together in a way that lies beyond the classical worldview.

\section{Conclusions}

We began with the issue of compatibility between probability
assignments.  Extending these considerations from the classical realm
to the quantum, we found that the problem of PP-ODOP compatibility for
three pure states in three-dimensional Hilbert space leads naturally
to SICs and MUB.  There are still open questions regarding PP
compatibility in higher dimensions~\cite{Brun2015}, and SIC solutions in
higher dimensions grow much more complicated than
Eq.~(\ref{eq:ME-SIC}). However, the patterns of linear dependencies
observed in higher-dimensional SICs~\cite{Dang2013} suggest that SIC
states may have interesting compatibility properties there as well.
Likewise, the strategy of finding the pure states whose SIC
representations have minimal Shannon entropy yields an intriguing
result in dimension 8~\cite{Szymusiak2015}, and perhaps such states
merit attention more generally.

\pagebreak

I thank Marcus Appleby, Chris Fuchs and R\"udiger Schack for
illuminating discussions.  I also thank the anonymous referees for
helpful feedback.  At the time this paper was first drafted, I was
supported by the Brandeis Geometry and Dynamics IGERT grant
(number NSF-DGE1068620).


\begin{thebibliography}{999}

\bibitem{appleby2009} D.\ M.\ Appleby, ``SIC-POVMs and
  MUBs: Geometrical relationships in prime dimension,''
  \arxiv[quant-ph]{0905.1428} (2009).

\bibitem{appleby2011} D.\ M.\ Appleby, \AA.\ Ericsson and
  C.\ A.\ Fuchs,
  ``\hrefdoi{10.1007/s10701-010-9458-7}{Properties of
    QBist State Spaces},'' \booktitle{Foundations of Physics} {\bf
    41,} 3 (2011), 564--79, \arxiv[quant-ph]{0910.2750}.

\bibitem{Bengtsson10} I.\ Bengtsson, K.\ Blanchfield and A.\ Cabello,
  ``\hrefdoi{10.1016/j.physleta.2011.12.011}{A
  Kochen--Specker inequality from a SIC},'' \booktitle{Physics Letters
  A} {\bf 376,} 4 (2012), 374--76, \arxiv[quant-ph]{1109.6514}.

\bibitem{Brun2015} T.\ A.\ Brun, M.-H.\ Hsieh and C.\ Perry,
  ``\hrefdoi{10.1103/PhysRevA.92.012107}{Compatibility
  of state assignments and pooling of information},''
  \booktitle{Physical Review A} \textbf{92,} 1 (2015), 012107,
  \arxiv[quant-ph]{1310.5325}.

\bibitem{Cabello2011} A.\ Cabello, ``State-independent quantum
  contextuality and maximum nonlocality,'' \arxiv[quant-ph]{1112.5149}
  (2011).

\bibitem{Cabello2015} A.\ Cabello, M.\ Kleinmann and C.\ Budroni,
  ``\hrefdoi{10.1103/PhysRevLett.114.250402}{Necessary
  and sufficient condition for quantum state-independent
  contextuality},'' \booktitle{Physical Review Letters} {\bf 114,} 25
  (2015), 250402, \arxiv[quant-ph]{1501.03432}.

\bibitem{Cameron2002} P.\ J.\ Cameron, ``Steiner
  Triple Systems,'' \booktitle{Encyclop{\ae}dia of Design Theory}
  (2002), \url{http://designtheory.org/library/encyc/sts/g/}.

\bibitem{CavesFuchsSchack2002} C.\ M.\ Caves, C.\ A.\ Fuchs and
  R.\ Schack,
  ``\hrefdoi{10.1103/PhysRevA.66.062111}{Conditions
    for compatibility of quantum state assignments},''
  \booktitle{Physical Review A} {\bf 66,} 6 (2002), 062111,
  \arxiv{quant-ph/0206110}.

\bibitem{combes2015} J.\ Combes, C.\ Ferrie, M.\ S.\ Leifer and
  M.\ F.\ Pusey, ``Why protective measurement does not establish the
  reality of the quantum state,'' \arxiv[quant-ph]{1509.08893} (2015).

\bibitem{Dang2013} H.\ B.\ Dang, K.\ Blanchfield, I.\ Bengtsson and
  D.\ M.\ Appleby,
  ``\hrefdoi{10.1007/s11128-013-0609-6}{Linear
    dependencies in Weyl--Heisenberg orbits},'' \booktitle{Quantum
    Information Processing} {\bf 12,} 11 (2013), 3449--75,
  \arxiv[quant-ph]{1211.0215}.


\bibitem{Voldemort} C.\ A.\ Fuchs, ``QBism: The Perimeter of Quantum
  Bayesianism,'' \\ \arxiv[quant-ph]{1003.5209} (2010).

\bibitem{RMP} C.\ A.\ Fuchs and R.\ Schack,
  ``\hrefdoi{10.1103/RevModPhys.85.1693}{Quantum-Bayesian
  Coherence},'' \booktitle{Reviews of Modern Physics} {\bf 85,} 4
  (2013), 1693--1715, \arxiv[quant-ph]{1301.3274}.

\bibitem{Fuchs2014} C.\ A.\ Fuchs and B.\ C.\ Stacey,
  ``Some negative remarks on operational approaches to quantum
  theory.'' In \booktitle{Quantum Theory: Informational Foundations
    and Foils} (Springer, 2015). \arxiv[quant-ph]{1401.7254}.

\bibitem{Gibbons2004} K.\ Gibbons, M.\ J. Hoffman and
  W.\ K.\ Wootters,
  ``\hrefdoi{10.1103/PhysRevA.70.062101}{Discrete phase
    space based on finite fields},'' \booktitle{Physical Review A}
  \textbf{70,} 6 (2004), 062101, \arxiv{quant-ph/0401155}.

\bibitem{ConicalDesigns} M.\ A.\ Graydon and D.\ M.\ Appleby,
  ``\hrefdoi{10.1088/1751-8113/49/8/085301}{Quantum
  conical designs},'' \booktitle{Journal of Physics A} \textbf{49,} 8
  (2016), 085301, \arxiv[quant-ph]{1507.05323}.

\bibitem{mermin1993} N.\ D.\ Mermin,
  ``\href{http://dx.doi.org/10.1103/RevModPhys.65.803}{Hidden
  variables and the two theorems of {John} {Bell}},''
  \booktitle{Reviews of Modern Physics} {\bf 65,} 3 (1993), 803--15.

\bibitem{PBR} M.\ F.\ Pusey, J.\ Barrett and T.\ Rudolph,
  ``\hrefdoi{10.1038/nphys2309}{On the reality of the
  quantum state},'' \booktitle{Nature Physics} {\bf 8} (2012),
  475--78, \arxiv[quant-ph]{1111.3328}.

\bibitem{Renes04} J.\ M.\ Renes, R.\ Blume-Kohout, A.\ J.\ Scott and
  C.\ M.\ Caves,
  ``\hrefdoi{10.1063/1.1737053}{Symmetric
    informationally complete quantum measurements},''
  \booktitle{Journal of Mathematical Physics} {\bf 45,} 6 (2004),
  2171, \arxiv{quant-ph/0310075}.

\bibitem{Schlosshauer14} M.\ Schlosshauer and A.\ Fine,
  ``\hrefdoi{10.1103/PhysRevLett.112.070407}{No-go
  theorem for the composition of quantum systems},''
  \booktitle{Physical Review Letters} {\bf 112,} 7 (2014), 070407,
  \arxiv[quant-ph]{1306.5805}.

\bibitem{schwinger1960} J.\ Schwinger, ``Unitary
  operator bases,'' \booktitle{PNAS} {\bf 46} (1960), 570--79,
  \url{http://www.pnas.org/content/46/4/570}.

\bibitem{Scott10} A.\ J.\ Scott and M.\ Grassl,
  ``\hrefdoi{10.1063/1.3374022}{SIC-POVMs:\ A new
  computer study},'' \booktitle{Journal of Mathematical Physics} {\bf
  51,} 4 (2010), 042203, \arxiv[quant-ph]{0910.5784}.

\bibitem{spekkens2007} R.\ W.\ Spekkens,
  ``\href{http://dx.doi.org/10.1103/PhysRevA.75.032110}{Evidence for
  the epistemic view of quantum states: A toy theory},''
  \booktitle{Physical Review A} {\bf 75,} 3 (2007) 032110,
  \arxiv{quant-ph/0401052}.

\bibitem{bartlett2012} S.\ D.\ Bartlett, T.\ Rudolph, and
  R.\ W.\ Spekkens,
  ``\href{http://dx.doi.org/10.1103/PhysRevA.86.012103}{Reconstruction
    of {Gaussian} quantum mechanics from {Liouville} mechanics with an
    epistemic restriction},'' \booktitle{Physical Review A} {\bf 86,}
  1 (2012), 012103, \arxiv[quant-ph]{1111.5057}.

\bibitem{spekkens2014} R.\ W.\ Spekkens, ``Quasi-quantization:
  classical statistical theories with an epistemic restriction.''  In
  \booktitle{Quantum Theory: Informational Foundations and Foils}
  (Springer, 2015).  \arxiv[quant-ph]{1409.5041}.

\bibitem{stacey-thesis} B.\ C.\ Stacey, {\sl Multiscale Structure in
  Eco-Evolutionary Dynamics.} PhD thesis, Brandeis University, 2015.
  \arxiv[q-bio.PE]{1509.02958}.

\bibitem{Szymusiak2015} A.\ Szymusiak and W.\ S\l{}omczy\'nski,
  ``Informational power of the Hoggar SIC-POVM,''
  \arxiv[quant-ph]{1512.01735} (2015).

\bibitem{Tabia12} G.\ N.\ M.\ Tabia,
  ``\hrefdoi{10.1103/PhysRevA.86.062107}{Experimental
  scheme for qubit and qutrit symmetric informationally complete
  positive operator-valued measurements using multiport devices},''
  \booktitle{Physical Review A} {\bf 86,} 6 (2012), 062107,
  \arxiv[quant-ph]{1207.6035}.

\bibitem{TabiaAppleby13} G.\ N.\ M.\ Tabia and D.\ M.\ Appleby,
  ``\hrefdoi{10.1103/PhysRevA.88.012131}{Exploring the
  geometry of qutrit state space using symmetric informationally
  complete probabilities},'' \booktitle{Physical Review A} {\bf 88,} 1
  (2013), 012131, \arxiv[quant-ph]{1304.8075}.

\bibitem{vanenk2007} S.\ J.\ van Enk,
  ``\href{http://dx.doi.org/10.1007/s10701-007-9171-3}{A toy model for
  quantum mechanics},'' \booktitle{Foundations of Physics} {\bf 37,} 10
  (2007) 1447--60, \arxiv[quant-ph]{0705.2742}.

\bibitem{Veitch14} V.\ Veitch, S.\ A.\ H.\ Mousavian, D.\ Gottesman
  and J.\ Emerson,
  ``\hrefdoi{10.1088/1367-2630/16/1/013009}{The
    resource theory of stabilizer computation},'' \booktitle{New
    Journal of Physics} {\bf 16} (2014), 013009,
  \arxiv[quant-ph]{1307.7171}.

\bibitem{Wootters1987} W.\ K.\ Wootters,
  ``\hrefdoi{10.1016/0003-4916(87)90176-X}{A
  Wigner-Function Formulation of Finite-State Quantum Mechanics},''
  \booktitle{Annals of Physics} {\bf 176} (1987), 1--21.

\bibitem{Zauner99} G. Zauner, \booktitle{Quantum Designs --
  Foundations of a Noncommutative Theory of Designs.} PhD thesis,
  University of Vienna, 1999. \url{http://www.gerhardzauner.at/qdmye.html}.

\bibitem{Zhu2014} H.\ Zhu,
  ``\hrefdoi{10.1016/j.aop.2015.08.005}{Super-symmetric
  informationally complete measurements},'' \booktitle{Annals of
  Physics} {\bf 362} (2015), 311--26, \arxiv[quant-ph]{1412.1099}.

\bibitem{Zhu2016} H.\ Zhu, ``Quasiprobability representations of
  quantum mechanics with minimal negativity,''
  \arxiv[quant-ph]{1604.06974} (2016).

\end{thebibliography}
\end{document}